\documentclass{ws-rv9x6}
\usepackage{subfigure}   % required only when side-by-side / subfigures are used
\usepackage{ws-rv-thm}   % comment this line when `amsthm / theorem / ntheorem` package is used
\usepackage[square]{ws-rv-van}
\usepackage{enumitem}   % numbered citation & references (default)
\makeindex
\newcommand{\hs}{\hspace{1mm}}
\newcommand{\nar}{New Astronomy Reviews}

\def\gsim{~\rlap{$>$}{\lower 1.0ex\hbox{$\sim$}}}
\def\lsim{~\rlap{$<$}{\lower 1.0ex\hbox{$\sim$}}}

\def\aj{AJ}%
\def\araa{ARA\&A}%
\def\apj{ApJ}%
\def\pasa{PASA}

\def\aap{A\&A}%
\def\aapr{A\&A~Rev.}%
\def\mnras{MNRAS}%
\def\prd{Phys.~Rev.~D}%
\def\pasp{PASP}%
\def\sovast{Soviet~Ast.}%
\def\ssr{Space~Sci.~Rev.}%
\def\nat{Nature}%
\def\physrep{Phys.~Rep.}%
\begin{document}

\setcounter{chapter}{13}
\title{Formation of the First Black Holes}
\chapter[Prospects for Detecting the First BHs with the Next Generation of Telescopes]{Prospects for Detecting the First Black Holes with the Next Generation of Telescopes}\label{chapter:dijkstra}\footnotetext{$^1$ Preprint~of~a~review volume chapter to be published in Latif, M., \& Schleicher, D.R.G., `Prospects for Detecting the First Black Holes with the Next Generation of Telescopes', Formation of the First Black Holes, 2018 \textcopyright Copyright World Scientific Publishing Company, \url{https://www.worldscientific.com/worldscibooks/10.1142/10652}}
\author[Dijkstra]{Mark Dijkstra}

\address{Institute of Theoretical Astrophysics, University of Oslo,\\
P.O. Box 1029 Blindern, NO-0315 Oslo, Norway\\
astromark77@gmail.com\footnote{Currently at Stitch Fix Inc, 1, Montgomery tower, San Francisco, CA 94104, USA}}

\begin{abstract}
This chapter describes the prospects for detecting the first black holes in our Universe, with a specific focus on instruments/telescopes that will come online within the next decade, including e.g. the SKA, WFIRST, EUCLID, JWST, and large ground-based facilities such as E-ELTs, GMT and TMT. This chapter focusses on: ({\it i}) the indirect detectability of the first generations of stellar mass black holes through their imprint on various Cosmic Background Radiation fields including the 21-cm and X-ray backgrounds; ({\it ii}) direct detectability of line and continuum emission by more massive black holes ($M_{\rm BH} \sim 10^4-10^6 M_{\odot}$) that formed via channels other than `ordinary' stellar evolution. 
\end{abstract}
%\markright{Customized Running Head for Odd Page} % default is Chapter Title.
\body

\section{Introduction}\label{Intro}

Understanding the formation of the first stars, black holes and galaxies in our Universe represents one of the key challenges in cosmology. This represents a well-defined problem, with initial conditions that are constrained well by the observed intensity fluctuations in the cosmic microwave background (CMB). However, understanding the physical properties of the first stars and black holes requires following the evolution of density perturbations deep into the non-linear regime, where a number of physical processes - e.g. ({\it i}) the effects of turbulence on sub-grid scales, which can affect hydrodynamics on resolved scales \citep{Latif2013c}, ({\it ii}) the presence of weak seed magnetic fields \citep{Sethi10,VanBorm2013,LatifMag2014}, ({\it iii}) the precise nature of dark matter (cold vs. warm vs. self-interacting), ({\it iv}) streaming velocities between dark matter and baryons \citep{Tseliakhovich2010,Maio11,Oleary2012,Latif2014Stream}, and ({\it v}) feedback effects that become relevant with the first onset of nuclear fusion (as in `quasi-stars', see e.g. Ref~\citet{Begelman2008}) - can strongly affect the late stages of gravitational collapse (see e.g. \citet{Volonteri10,Haiman13,LatifFerrara2016} for reviews on the formation of the first black holes and \citet{Bromm04,Greif15} for the formation of the first stars). With little to no guidance by observations, the physical properties of the first stars and black holes are not well known.

The goal of this chapter is to discuss the prospects for improving our observational constraints, and to focus on the prospects for detecting the first black holes that formed in the Universe. I will focus mainly\footnote{I will not discuss the detectability of the first primordial black holes that may form in the very early Universe (see e.g. Refs.~\citet{Zeldovich66,Hawking}). This chapter will in particular be dedicated to electromagnetic observations, while observations through gravitational wave observatories have been discussed in chapter~3. For observational constraints on primordial black holes, see e.g. Ref~\citet{Mack08}.} on the detectability of ({\it i}) stellar mass black holes, which formed after at the end of the life-time of the first massive stars, and ({\it ii}) more massive $M_{\rm BH}\sim 10^4-10^6 M_{\odot}$ that can form in the young Universe through a variety of channels. This discussion focusses on the detectability for the most recent and the next generation of instruments and telescopes. The outline of this chapter is as follows: I will discuss the prospects for the detectability of stellar mass black holes in \S~\ref{sec:stellardirect} and \S~\ref{sec:stellarindirect}. I then discuss the detectability of more massive black holes in \S~\ref{sec:massive}, before I summarize in \S~\ref{sec:conc}. 

\begin{table}[htp]
\caption{default}
\begin{center}
\begin{tabular}{|c|c|}
Abbreviation & full name\\ \hline
BH & black hole\\
DCBH & direct collapse black hole\\
CMB & cosmic microwave background\\
IGM & intergalactic medium\\
WF & Wouthuysen-Field \\
XRB & X-ray background\\
HMXB & high-mass X-ray binary\\
SFR & star formation rate\\
SFG & star-forming galaxy\\
NIRB & near infrared background\\
OBG & obese black hole galaxy
\end{tabular}
\end{center}
\label{abrsuper}
\end{table}%

\section{Prospects for Direct Detection of the First Stellar Mass Black Holes}\label{sec:stellardirect}

Observations indicate that the electro-magnetic spectrum associated with accretion onto a stellar mass black hole varies strongly with the rate at which the black hole is accreting. For the highest accretion rate - $\dot{m}\gsim 10^{-2}\dot{m}_{\rm edd}$, where\footnote{For a radiative efficiency $\eta$, the maximum mass accretion rate equals $\dot{m}=L_{\rm edd}/(\eta c^2)$.} $\dot{m}_{\rm edd}\equiv L_{\rm edd}/c^2$ - the accreting black hole is in its high-soft state, during which the spectrum can be decomposed into two components\footnote{There is a third 'reflection' component, which consists of high-energy photons originating in the hot corona, but which scattered (either by free electrons, or fluorescently by ions) off the accretion disk. We follow Ref.~\refcite{Yue13} and assume this component to be sub-dominant to the other two.}: ({\it i}) at low energies the standard Shakura-Sunyaev disk model (see e.g. Refs.~\refcite{Shakura,Shimura15}) applies, which can be approximated as $f_{\nu} \propto \nu^{1/3}$ out to $h_{\rm p}\nu \lsim kT_{\rm max}\sim 1m_{\rm BH}^{-1/4}$ keV, where $m_{\rm BH}$ denotes the black hole mass (in $M_{\odot}$); ({\it ii}) a power-law component $f_{\nu} \propto \nu^{-\alpha_{X}}$ which extends out to $\sim$ a few to a few hundreds keV (after which it cuts off exponentially, see e.g. Refs \refcite{Sazonov04,Feng11}), with $\alpha_X$ having a broad distribution covering $0 \lsim \alpha_X \lsim 2$, and peaked at $\alpha_X \sim 1$ (e.g. Ref~\refcite{Feng11}). This power-law component is thought to originate in a hot corona of gas surrounding the black hole which Compton scatters UV photons emitted by the disk to higher energies. The fraction of the total bolometric luminosity that is in either component of the spectrum varies: the power-law component can contain $\sim$ a few to a few tens of per cent of the total bolometric luminosity (e.g. Ref~\refcite{Gilfanov14}). In other words, the disk component typically contains a fraction of the bolometric luminosity which is of order unity. For a black hole of mass $m_{\rm BH}$ accreting at its Eddington limit, the flux density at an observed frequency $\nu_{\rm obs}$ can be estimated from $f_\nu \sim (1+z)L_{\rm bol}/[4 \pi d^2_{\rm L}(z)\nu_{\rm max}]$ (where $d_{\rm L}(z)$ denotes the luminosity distance to redshift $z$). Plugging in some numerical values yields
\begin{equation}
f_\nu \approx 2 \times 10^{-36}\left (\frac{m_{\rm BH}}{100M_{\odot}}\right)^{5/4}\left(\frac{h_{\rm p}\nu_{\rm obs}}{10\hspace{1mm}{\rm eV}}\right)^{1/3}\left( \frac{1+z}{10}\right)^{-1.5}\hspace{1mm}\frac{{\rm erg}}{{\rm s}\hspace{1mm}{\rm cm}^2\hspace{1mm}{\rm Hz}},
\end{equation} for $h_{\rm p}\nu_{\rm obs} \lsim kT_{\rm max}/(1+z)$, where used that $d^2_{\rm L}(z)\propto (1+z)^{2.5}$ at $z\gg1$ (approximately). The $m_{\rm BH}^{5/4}$-dependence arises because $\nu_{\rm max} \propto m_{\rm BH}^{-1/4}$. This flux density corresponds to an AB-magnitude of
\begin{equation}\label{eq:mab}
m_{\rm AB}\sim 41-3.1\log\left(\frac{m_{\rm BH}}{100M_{\odot}}\right)-0.83\log\left(\frac{\nu}{10\hspace{1mm}{\rm eV}}\right)+3.8 \log \left( \frac{1+z}{10}\right),
\end{equation} for $h_{\rm p}\nu_{\rm obs} \lsim kT_{\rm max}/(1+z)$.  For comparison, NIRCAM on the the soon-to-be launched James Webb Space Telescope (JWST) can detect a point source with $S/N=10$ in after an integration of $t=10^4$ with an AB-magnitude of $\sim 29$ in the observed wavelength range $\lambda=1-5 \mu$m\footnote{\url{https://jwst.stsci.edu/science-planning/proposal-planning-toolbox/sensitivity-overview}}. 
Larger ground-based telescopes (such as E-ELT) may go $\sim$ 3 magnitudes deeper, but which is not nearly enough to detect continuum emission from the first black holes accreting at Eddington luminosity.\\

\section{Prospects for Indirect Detection of the First Stellar Mass Black Holes}\label{sec:stellarindirect}

\subsection{Detecting the First BH through 21-cm}\label{sec:21cm}
The prospects for {\it indirect} detection of the first black holes are much better. X-rays produced in the inner disk or hot corona can penetrate deeply into the neutral intergalactic medium where a significant fraction of their energy is converted into heat\footnote{In a fully neutral IGM, photons with $E\gsim 50$ eV deposit $\sim 1/3$ of their energy as heat, $\sim 1/3$ goes into ionization, and the remaining $\sim 1/3$ goes into collisional excitation of transitions in H and He/He$^+$ (e.g. Refs~\refcite{Shull85,Valdes10}). As the ionization fraction rises above $\sim 10\%$ however, the fraction of photon energy that goes into heat rapidly approaches unity.}. This re-heating of the IGM has a direct observational implication for the detectability of the 21-cm hyperfine transition of atomic hydrogen in the ground state (e.g. Ref~\refcite{Mirabel11}). Since this line provides such a potentially powerful probe of the high-z Universe (and the first black holes), we summarize the basic physics that regulates its detectability in a bit more detail below. \\

The visibility of HI gas in its 21-cm line -  with frequency $\nu_{21}\sim 1.4$ Ghz - is expressed as a {\it differential} brightness temperature with respect to the background Cosmic Microwave Background (CMB), $\delta T_{\rm b}(\nu)$ \cite{Furlanetto06,Morales10}

\begin{equation}\label{eq:tb}
\delta T_{\rm b}(\nu) \approx 9x_{\rm HI}(1+\delta)(1+z)^{1/2}\left( 1-\frac{T_{\rm CMB}(z)}{T_{\rm S}} \right) \left[ \frac{H(z)(1+z)}{dv_{||}/dr} \right] \hs{\rm mK},
\end{equation} where $\delta+1 \equiv \rho/\bar{\rho}$ denotes the overdensity of the gas, $z$($=[\nu_{21}/\nu]-1$) the redshift of the gas, $T_{\rm CMB}(z)=2.73(1+z)$ K denotes the temperature of the CMB, the factor in square brackets contains the line-of-sight velocity gradient $dv_{||}/dr$. The term $T_{\rm S }$ denotes the spin (or excitation) temperature of the 21-cm transition, which quantifies the number densities of hydrogen atoms in each of the hyperfine transitions, and which is discussed in more detail below (see Eq~\ref{eq:ts}). The numerical prefactor of $9$ mK in Eq~(\ref{eq:tb}) applies for gas undergoing Hubble expansion. For slower expansion rates, we increase the number of hydrogen atoms within a fixed velocity (and therefore frequency) range, which enhances the brightness temperature. \\

Eq~\ref{eq:tb} states that when $T_{\rm CMB}(z) < T_{\rm S}$, $\delta T_{\rm b}(\nu) > 0$, and vice versa. This means that we see HI in absorption [emission] when $T_{\rm S}< T_{\rm CMB}(z)$ [$T_{\rm S}> T_{\rm CMB}(z)$]. The spin temperature thus plays a key role in setting the 21-cm signal, and we briefly discuss the physical processes that determine $T_{\rm s}$ below. Eq~\ref{eq:tb} highlights why so much effort is going into trying to detect the 21-cm line: having a 3D map of $\delta T_{\rm b}(\nu)$ provides us with direct constraints on the 3D density [$\delta({\bf r})$], and line-of-sight velocity field [$v_{||}({\bf r})$] of baryons in our Universe. It therefore provides us with a powerful cosmological and astrophysical probe of the high-redshift Universe.\\

\begin{figure}
\centering
\vspace{-5mm}
\begin{minipage}{.5\textwidth}
\centering
\includegraphics[width=6.0cm,angle=0]{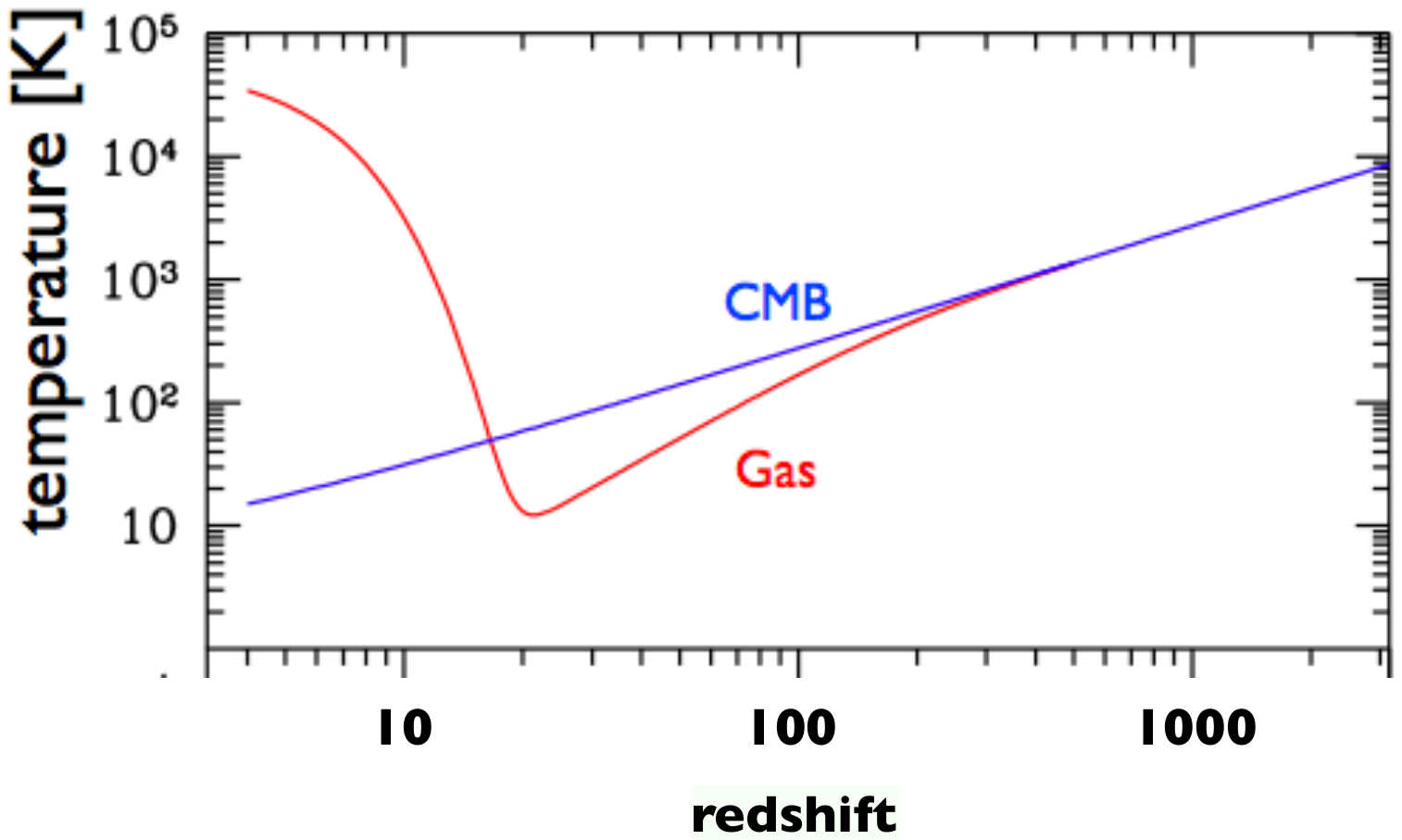}
\end{minipage}%
\begin{minipage}{.5\textwidth}
\centering
\includegraphics[width=6.0cm,angle=0]{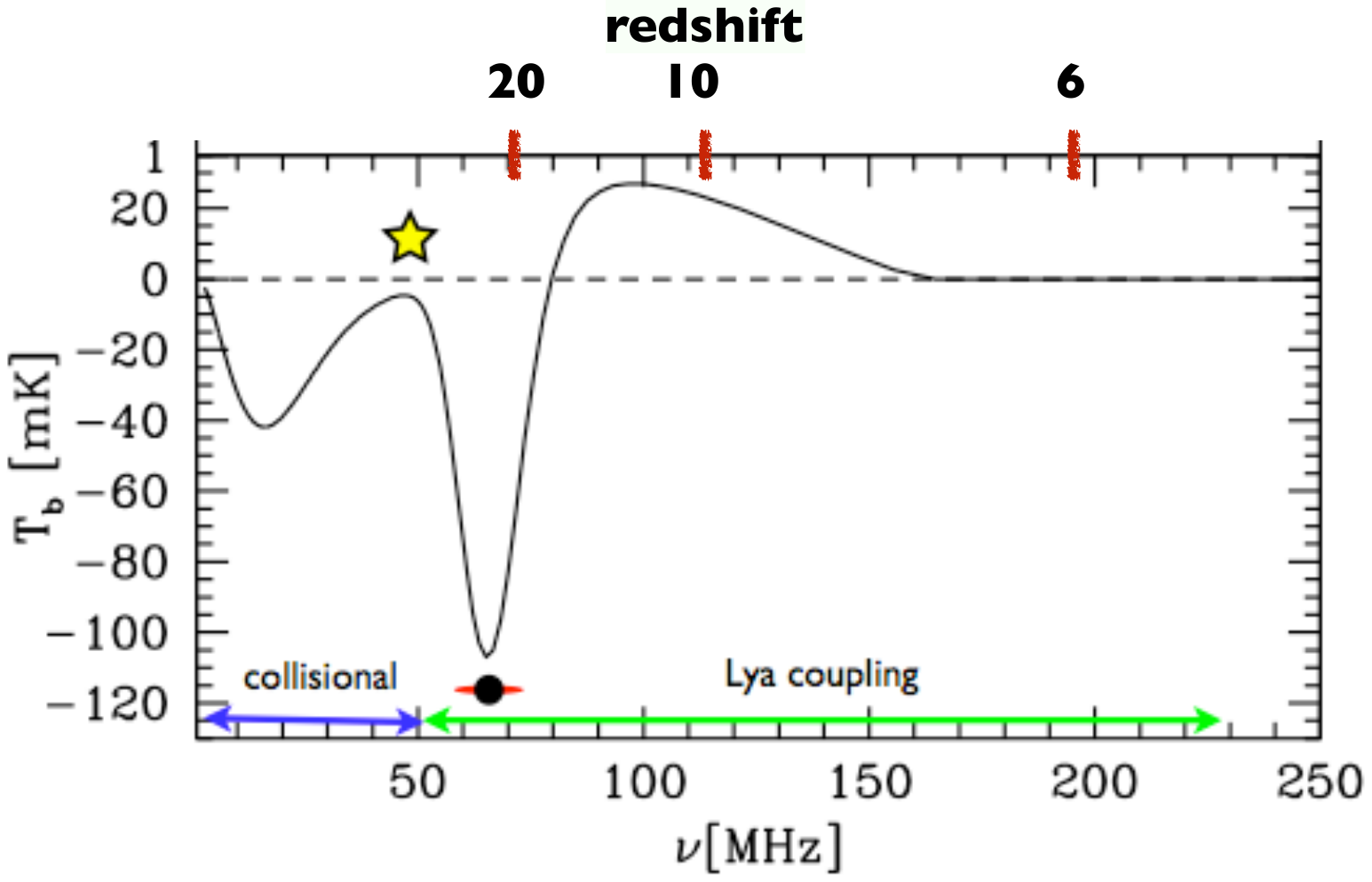}
\end{minipage}%
\vspace{1mm}
\caption[]{{\it Left panel:} Volume-averaged redshift evolution of $T_{\rm gas}$ ({\it red solid line}) and the $T_{\rm CMB}$ ({\it blue solid line}).  {Right panel:} The `Global' 21-signal, represents the sky-averaged 21-cm brightness temperature $\delta T_{\rm b}(\nu)$. See the main text for a description of each of these curves. The global 21-cm signal constrains when the first stars, black holes, and galaxies formed and depends on their spectra. {\it Credit: adapted from slides created by J. Pritchard (based on Refs~\refcite{PL10,PL12}}).}
\label{fig:global}
\end{figure}

The spin temperature $T_{\rm S}$ is set by ({\it i}) collisions, which drive $T_{\rm S} \rightarrow T_{\rm gas}$, ({\it ii}) absorption by CMB photons, which drives $T_{\rm S} \rightarrow T_{\rm CMB}$ (and thus that $\delta T_{\rm b} \rightarrow 0$, see Eq~\ref{eq:tb}), and ({\it iii}) Ly$\alpha$ scattering. Ly$\alpha$ scattering mixes the two hyperfine levels, which drive $T_{\rm S} \rightarrow T_{\rm gas}$, and effect which is known as the `Wouthuysen-Field effect' (WF effect). Quantitatively, it has been shown that\cite{Madau97,Furlanetto06,Morales10}
\begin{equation}\label{eq:ts}
\frac{1}{T_{\rm S}}=\frac{T^{-1}_{\rm CMB}+x_c T^{-1}_{\rm gas}+x_{\alpha}T^{-1}_{\rm gas}}{1+x_c+x_{\alpha}},
\end{equation} where $x_{c}=\frac{C_{21}T_*}{A_{21}T_{\rm gas}}$ denotes the collisional coupling coefficient, in which $k_{\rm B}T_*$ denotes the energy difference between the hyperfine levels, $C_{21}$ denotes the collisional de-excitation rate coefficient, and $A_{21}=2.85\times 10^{-15}$ s$^{-1}$ denotes the Einstein-A coefficient for the 21 cm
transition. Furthermore, $x_{\alpha}=\frac{4P_{\alpha}}{27A_{21}T_{\rm gas}}$, in which $P_{\alpha}$ denotes the Ly$\alpha$ scattering rate\footnote{We have adopted the common assumption here that the Ly$\alpha$ color temperature $T_{\alpha}=T_{\rm gas}$.}. Eq~\ref{eq:ts} implies that the spin temperature $T_{\rm s}$ is a weighted average of the gas and CMB temperature. The {\it left panel} of Figure~\ref{fig:global} shows the universally (or globally) averaged temperatures of the gas ({\it red solid line}) and the CMB ({\it blue solid line}), and the corresponding `global' 21-cm signature in the {\it Right panel}. We discuss these below (this discussion follows Ref~\refcite{Saas}):\\

\begin{itemize}[leftmargin=*]
\item Adiabatic expansion causes $T_{\rm CMB} \propto (1+z)$ at all $z$. At $z \gsim 100$, $T_{\rm gas} \sim T_{\rm CMB}$ as a result of the interaction between CMB photons and the small residual fraction of free electrons that exist following recombination. When $T_{\rm CMB}=T_{\rm gas}$, we must have $T_{\rm s}=T_{\rm CMB}$, and therefore that $\delta T_{\rm b}(\nu)=0$ mK, which corresponds to the high-$z$ limit in the {\it right panel}.

\item At $z \lsim 100$, electron scattering can no longer couple the CMB and gas temperatures, and the (non-relativistic) baryons adiabatically cool as $T_{\rm gas} \propto (1+z)^2$. Because $T_{\rm gas} < T_{\rm CMB}$, we must have that $T_{\rm s} < T_{\rm CMB}$ and we expect to see the 21-cm line in absorption. When $T_{\rm gas}$ first decouples from $T_{\rm CMB}$ the gas densities are high enough for collisions to keep $T_{\rm s}$ locked to $T_{\rm gas}$. However, at $z\sim 70$ ($\nu \sim 20$ MHz) collisions can no longer couple $T_{\rm s}$ to $T_{\rm gas}$, and $T_{\rm s}$ crawls back to $T_{\rm CMB}$, which reduces $\delta T_{\rm b}(\nu)$ (at $\nu\sim 20-50$ MHz, i.e. $z\sim70-30$).

\item The first stars, galaxies, and accreting black holes emitted UV photons in the range $E=10.2-13.6$ eV. These photons travel freely through the neutral IGM, until they redshift into one of the Lyman series resonances, at which point a radiative cascade can produce Ly$\alpha$. The formation of the first stars thus generates a Ly$\alpha$ background, which initiates the WF-coupling, which forces $T_{\rm s}\rightarrow T_{\rm gas}$. The onset of Ly$\alpha$ scattering - and thus the WF coupling - causes $\delta T_{\rm b}(\nu)$ to drop sharply at $\nu \gsim 50$ Mhz ($z \lsim 30$).

\item At some point the X-rays produced by accreting black holes easily penetrate deep into the cold, neutral IGM, and reheat the gas. The {\it left panel} thus has $T_{\rm gas}$ increase at $z\sim 20$, which corresponds to onset of X-ray heating. In the {\it right panel} this onset occurs a bit earlier. This difference reflects that the redshift of all the features (minima and maxima) in $\delta T_{\rm b}(\nu)$ are model dependent, and not well-known (more on this below). The onset of X-ray heating (combined with increasingly efficient WF coupling to the build-up of the Ly$\alpha$ background) enhances $\delta T_{\rm b}(\nu)$ until it becomes positive when $T_{\rm gas} > T_{\rm CMB}$.

\item Finally, $\delta T_{\rm b}(\nu)$ reaches yet another maximum, which reflects that neutral, X-ray heated gas is reionized away by the ionizing UV-photons emitted by star forming galaxies and quasars. When reionization is complete, there is no diffuse intergalactic neutral hydrogen left, and $\delta T_{\rm b}(\nu) \rightarrow 0$. 
\end{itemize}

In detail, the onset and redshift evolution of Ly$\alpha$ coupling, X-ray heating, and reionization depend on the redshift evolution of the number densities of galaxies, and their spectral characteristics. All these are uncertain, and it is not possible to make robust predictions for the precise shape of the global 21-cm signature. Instead, one of the main challenges for observational cosmology is to measure the global 21-cm signal, and from this constrain the abundances and characteristic of first generations of galaxies in our Universe. Detecting the global 21-cm is challenging, but especially the deep absorption trough that is expected to exist just prior to the onset of X-ray heating - {\it which is tightly linked to the formation of the first black holes in our Universe} - at $\nu \sim 70$ MHz is something that may be detectable because of its characteristic spectral shape. Constraints on the global 21-cm signal already rule out scenarios in which the Universe goes from fully neutral at fully ionized over $\Delta z \lsim 0.06$ at $6 < z < 13$ (see Ref~\refcite{Bowman10}). While such rapid reionization scenarios are not considered physically plausible, it is remarkable that global 21-cm experiments already have the sensitivity to detect (rapid) changes in the ionization state of the IGM. Encouraged by this success, a number of other experiments aim to detect the global 21-cm signal, including e.g. EDGES 2\cite{Monsalve17}, LEDA\footnote{\url{http://www.tauceti.caltech.edu/leda/}}, BIGHORNS\cite{Soko15}, and SARAS 2\cite{Saras2}.\\

\begin{figure}
\centerline{\includegraphics[width=11cm]{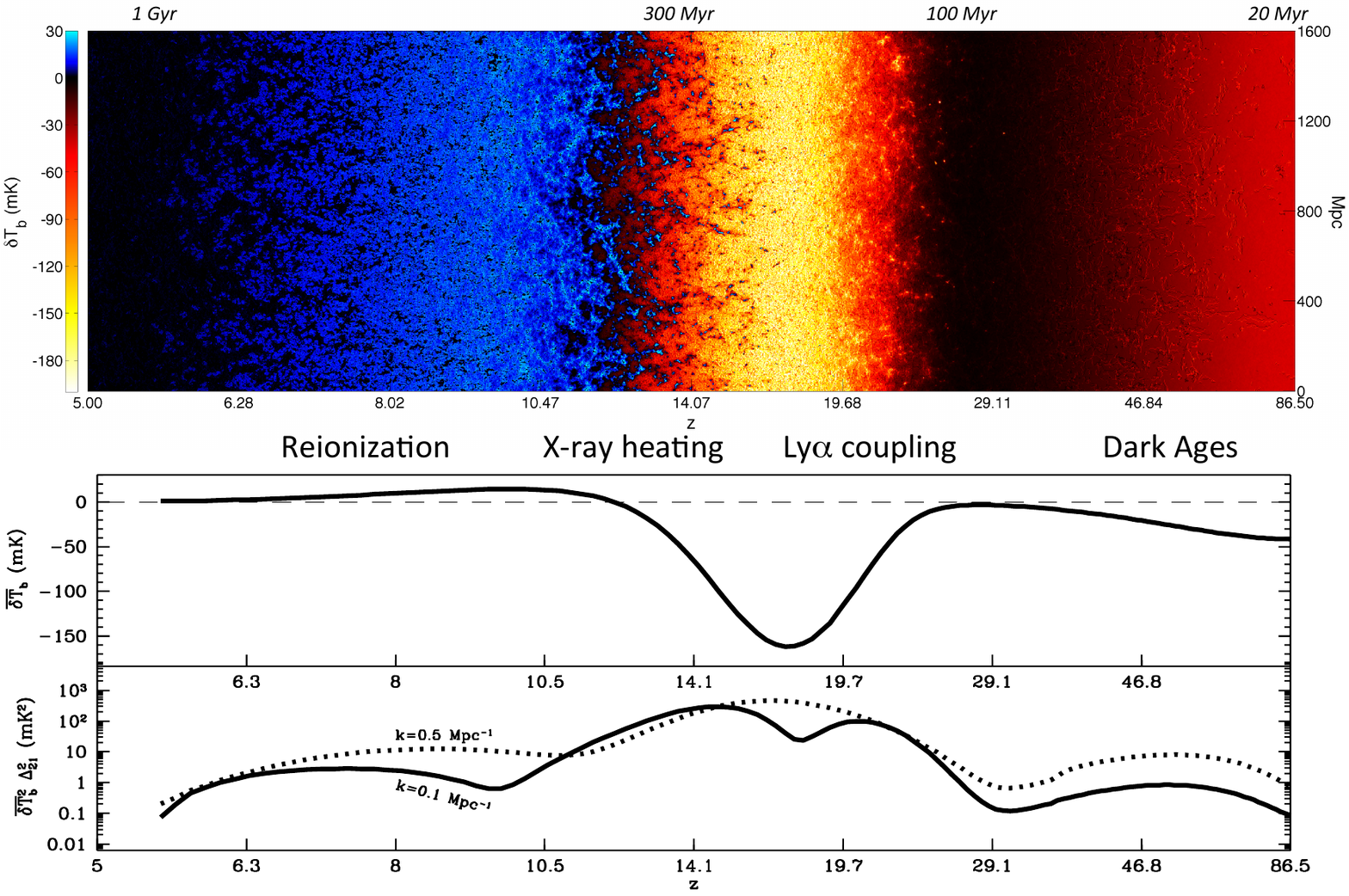}}
\caption{This Figure illustrates that the spatial fluctuations of the 21-cm background provide additional constraints on the first black holes in our Universe (see text) {\it Credit: Adopted from Figure 4 of Ref~\citet{Mesinger16}, `The Evolution Of 21 cm Structure (EOS): public, large-scale simulations of Cosmic Dawn and reionization', MNRAS, 459, 2342M. Reproduced by permission of Oxford University Press / on behalf of the RAS.}}
\label{fig:xray}
\end{figure}

Figure~\ref{fig:xray} illustrates that there is more information on the first black holes encoded within 21-cm observations: while the global 21-cm signal assigns a single, volume averaged, brightness temperature to each redshift, the {\it top panel} shows $\delta T_{\rm b}(\nu)$ through a slice of the Universe, which shows that at fixed $z$ there exist spatial fluctuations in $\delta T_{\rm b}(\nu)$. The {\it bottom panel} shows that the redshift dependence of the {\it variance} $\delta T^2_{\rm b}$ depends on scale (here represented with wavenumber $k$). The 3D power-spectrum of the 21-cm fluctuations (which corresponds to $\sim \delta T^2_{\rm b}(k)$) therefore contains additional constraints on the first black holes in the Universe\cite{Pritchard07,Mesinger13,Fialkov14,Mesinger16}.\\

There is a number of `first generation' (e.g. PAPER\cite{Parsons10}, LOFAR\cite{LOFAR}, MWA\cite{MWA}) and future (e.g. SKA\footnote{\url{http://skatelescope.org/}}, HERA\cite{HERA}) low frequency interferometers that aim to detect fluctuations in the 21-cm background on the scales shown in Figure~\ref{fig:xray}. Ref. \refcite{Mesinger16} finds that SKA1-low and HERA should be able to detect the epochs of reionization and X-ray heating at a signal-to-noise level of $\sim$ hundreds. With optimistic assumptions on the foregrounds, they also find that all first generations of instruments could detect these epochs at $S/N \gsim 3$. In fact, upper limits of the 21-cm power spectrum obtained by PAPER\cite{Pober15} already rule out some extreme models, in which the neutral IGM is not heated at all by X-rays even when reionization is well underway. Even though these extreme models are not really physical, it is extremely encouraging that observations have started placing constraints on the temperature of the low-density, neutral intergalactic medium in the very young Universe.

\subsection{Contribution of the Early Generations of Accreting BHs to Cosmic Backgrounds}

As discussed previously, X-rays emitted by the first black holes play an important role in shaping the thermal history of our Universe. A fraction of these X-rays may be observable today:  the neutral IGM is optically thin to photons with energies $E \gsim E_{\rm max} = 1.8[(1+z)/15)]^{0.5}x^{1/3}_{\rm HI}$ keV\cite{Dijkstra04}, and these photons would redshift
without absorption and would be observed as a present-day soft X-ray background (soft XRB). It is indeed possible that early generations of accreting black holes contributed significantly to the {\it unresolved} component of the XRBs:

\begin{itemize}[leftmargin=*]
\item The unresolved soft XRB (here, observed energies $E=0.5-2$ keV) can rule out certain models in which reionization is dominated by accreting black holes \cite{Dijkstra04,Salvaterra05,McQuinn12}, though quantitatively these constraints depend on the spectral slope $\alpha_{\rm X}$ characterizing the X-ray emission from the hot corona\cite{Dijkstra04,McQuinn12}.

\item Ref~\refcite{Dijkstra12} (hereafter D12) applied the locally inferred relation between a galaxies star formation rate and X-ray luminosity (e.g. Refs~\refcite{Lehmer10,Mineo12})\footnote{This correlation - $L_{\rm X}=2.6 \times 10^{39} \times {\rm SFR}$ erg s$^{-1}$, where SFR is measured in units of $M_{\odot}$ yr$^{-1}$ and X-ray luminosity is measured in the energy-range $E=0.5-8.0$ keV \cite{Mineo12}-  is driven by High-Mass X-ray Binaries (HMXBs). In HXMBs are binaries in which a compact object (neutron star or black hole) accretes gas that is stripping off a massive O or B binary companion star. Because these stars are short-lived HMXBs closely trace ongoing star formation. }, and showed that star forming galaxies that are too faint to be detected as individual X-ray sources can fully account for the observed unresolved soft XRB (where `soft' refers to the observed range $E=1-2$ keV). Interestingly, there are observational indications that the total X-ray luminosity from HMXBs per unit SFR increases with redshift\cite{Basu13}, which implies that star forming galaxies could contribute even more\footnote{Specifically, D12 computed that star forming galaxies (SFGs) with SFR$>10^{-2} M_{\odot}$ yr$^{-1}$ whose X-ray flux falls below the detection limit of the 2 Ms Chandra Deep Field North\cite{Alex03}, contribute $\sim 2-3 \times 10^{-13}$ erg s$^{-1}$ cm$^{-2}$ deg$^{-2}$ in the 1-2 keV band. In a more recent analysis, Ref~\refcite{Cap17} (hereafter C17) constrain the contribution of SFGs that are detected in the CANDELS catalogues of the 10 Ms Chandra Deep Fields to the XRB to be $\sim 3 \times 10^{-14}$ erg s$^{-1}$ cm$^{-2}$ deg$^{-2}$, which is a factor of $\sim 7-10$ times lower. There are several reasons for this difference, all of which relate to the fact that C17 considered a subset of SFGs that were included in the D12 analysis: ({\it i}) the 10 Ms Chandra observations allow for {\it direct} detection of more SFGs, which can therefore not contribute to the {\it unresolved} XRB; ({\it ii}) D12 include all star forming galaxies with SFR$>10^{-2} M_{\odot}$ yr$^{-1}$, which corresponds to $M_{\rm UV} \sim -13$ (not corrected for dust), most of which would be too faint to make it into the CANDELS catalogues; ({\it iii}) in D12 the largest contribution from star forming galaxies to the XRB comes comes from sources at $z<2$. C17 specifically notes that it is possible for faint sources at these redshifts to contribute significantly to the XRB. }. 

\item A more recent analysis, Ref~\refcite{Cap17} (which we referred to as C17) show that there are indications that the spectrum of the unresolved component of the Cosmic X-ray Background (XRB) is harder and more obscured than the collective spectrum of resolved sources. 
C17 propose a scenario in which a high-z ($z\gsim 6-7$) population of direct collapse black holes (DCBHs: black holes that form following the collapse of a pristine gas cloud directly into a massive black hole with no intermediate star formation\cite{Bromm03,Volonteri10,Haiman13,Latifrev}) fully account for the the remaining unresolved XRB. This proposal is motivated by theoretical modelling which showed that high-z DCBHs can - under certain assumptions - explain the amplitude of the observed fluctuations in the source extracted Near Infrared Background (NIRB) and its cross-correlation with the XRB \cite{Yue13}. C17 further show that if the unresolved XRB is indeed fully generated by a population of high-z Compton thick accreting black holes, that then observational constraints on the mass density of black holes require that the Compton thick gas that is surrounding the accreting black holes must have gas metallicities\footnote{This is because lower metallicity gas is more transparent to high energy X-rays. As a result, a {\it fixed} observed flux in the XRB translates to a {\it higher} accretion rate in enriched gas\cite{Priya17}, which translates to a larger mass density in accreting black holes.} $Z \lsim 10^{-3}Z_{\odot}$.

\end{itemize}

This discussion highlights that it is possible that early generations of accreting black holes contribute significantly to the unresolved cosmic X-ray backgrounds. However, the nature of the black holes varies greatly between different models, ranging from HMXB powering X-ray emission in sundetected ultra-faint $z<2$ star forming galaxies to $z \gg 6$ newly formed direct collapse black holes. It is possible to distinguish between these two models with future observations as we discuss in the next section.

\section{Prospects for Detection of the First Massive Black Holes}\label{sec:massive}

As already mentioned, the first black holes may have been much more massive, with masses possibly as large as $M_{\rm BH}\sim 10^{6} M_{\odot}$ (see Refs\refcite{Volonteri10,Haiman13,Latifrev} for a review of a variety of mechanisms that may give rise to such massive black hole seed masses). These massive objects are more easy to detect. In \S~\ref{sec:cont} we compare estimates of the continuum flux density, in \S~\ref{sec:line} we estimate fluxes in some lines, and in \S~\ref{sec:formation} we look at observational signatures associated with the formation of these black holes.

\subsection{Direct Detection of the Continuum Emission from First Massive BHs}\label{sec:cont}
\begin{figure} 
\centerline{\includegraphics[width=12.5cm]{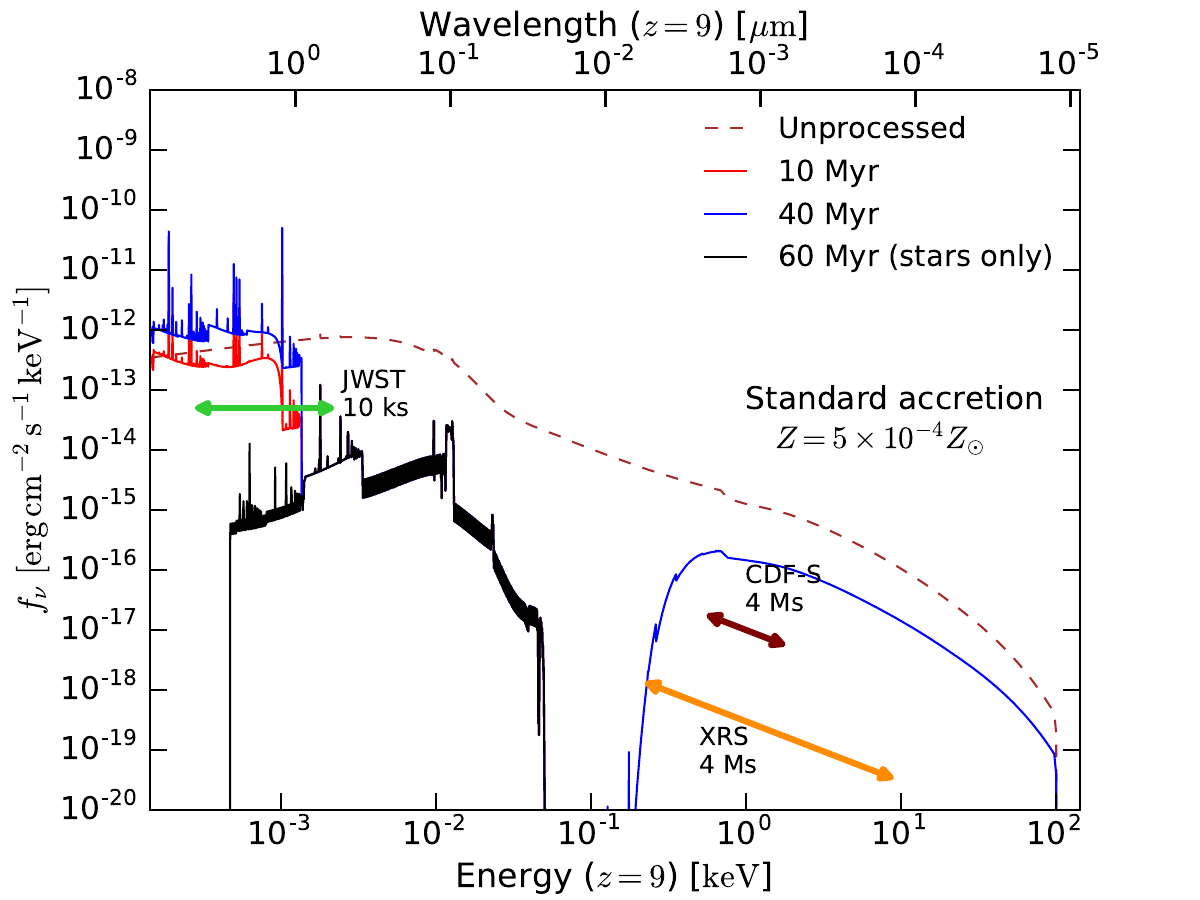}}
\caption{Model spectra of massive ($M_{\rm BH} \sim 10^4-10^6 M_{\odot}$) accreting black hole. The `unprocessed' spectrum consists of a disk + hot coronal component. Significant re-processing of ionizing radiation can occur by atomic hydrogen which is theoretically expected to be surrounding the accreting black hole. Photo-ionization and subsequent recombinating gives rise to strong nebular lines (Ly$\alpha$, H$\alpha$, He1640) which can be detected with future (and existing) telescopes \& instruments {\it Credit: from Figure 1 of Ref~\refcite{Priya17}. \textcopyright AAS. Reproduced with permission.}}
\label{fig:dcbhspec}
\end{figure}
The {\it red-dashed line} in Figure~\ref{fig:dcbhspec} shows an `unprocessed' (by HI surrounding the black hole, including intergalactic HI) model spectrum for a DCBH at $z\sim 9$ from Ref~\refcite{Priya17} for a number of assumptions: ({\it i}) the total X-ray flux from the accretion disk equals that coming from the hot corona; ({\it ii}) the spectrum coming from the corona is a power-law with slope $\alpha_{\rm X}=1$; ({\it iii}) the exponential cut-off of the spectrum occurs at high ($\gg 10$ keV) energies ({\it iv}) the DCBH has grown to $M_{\rm BH}=1.5\times 10^6 M_{\odot}$. The unprocessed flux density peaks at $h_{\rm p}\nu_{\rm obs}\sim kT_{\rm max}/(1+z)\sim 10$ eV at $f_{\nu} \sim 10^{-12.5}$ erg s$^{-1}$ cm$^{-2}$ keV$^{-1}$, which corresponds to $m_{\rm AB} \sim 26.2$. This agrees well with our estimate given in Eq~\ref{eq:mab}, which gives $m_{\rm AB}\sim 27$ for $m_{\rm AB}=1.5\times 10^6M_{\odot}$. \\

The {\it colored solid lines} show the spectrum after it has been processed through a large column density of HI gas (here $N_{\rm HI} \gsim 10^{24}$ cm$^{-2}$). This hydrogen gas absorbs most radiation blueward of the hydrogen ionization threshold at $h_{\rm p}\nu_{\rm obs}\gsim 13.6/(1+z)=1.4$ eV. Because the photoionization cross-section decreases $\propto \nu^{-3}$, the gas becomes increasingly transparent at higher energies, which causes the processes spectrum to rise at $h_{\rm p}\nu_{\rm obs} \gsim 0.2$ keV. Lines of different colors represent the accreting black hole at different times - and therefore black hole mass. Ionizing photons that are absorbed in the neutral gas are reprocessed into recombination lines (and continuum), which we will discuss in \S~\ref{sec:line}.\\

Figure~\ref{fig:dcbhspec} also indicates sensitivity of some existing and future observations: the {\it green line} shows the expected sensitivity of the James Webb Space Telescope in the near-IR ($\lambda \sim 1-5 \mu$m) for a $10^4$ s integration, which corresponds to $f_{\nu} \sim 10^{-13.3}$ erg s$^{-1}$ cm$^{-2}$ keV$^{-1}$ (i.e. $m_{\rm AB}\sim 28$, which agrees with the sensitivity estimate given in \S~\ref{sec:stellardirect}). Continuum magnitudes of $m_{\rm AB} \sim 26-27$ in the near-IR are also within reach of WFIRST (Wide Field Infrared Survey Telescope), which is a space-based NIR observatory whose 5$\sigma$ detection threshold for point sources is expected to reach $26.5-27.0$ at $\lambda\sim 1.0-1.7$ $\mu$m. Deep fields with EUCLID\cite{EUCLID} are expected to reach similar depth over $\sim 40$ deg$^{2}$. Finally, in `deep drilling fields' LSST can reach $m_{\rm AB} \sim 28$ though this is limited to $\lambda \lsim 1 \mu$m. Also shown are the sensitivity in the ({\it i}) existing 4Ms Chandra-Deep Field South, and ({\it ii}) the proposed `Lynx' X-ray telescope\footnote{\url{https://wwwastro.msfc.nasa.gov/lynx/}} which would have a sensitivity to point sources reaching $10^{-19}$ erg s$^{-1}$ cm$^{-2}$ - corresponding to a luminosity of $L_X\sim 10^{41}$ erg s$^{-1}$ at $z=10$- in a 4Ms observation\footnote{Another X-ray future telescope is ATHENA (\url{http://www.the-athena-x-ray-observatory.eu/mission.html}), which will reach a sensitivity of $10^{-17}$ erg s$^{-1}$ cm$^{-2}$, which can also detect the continuum of the more massive seeds ($M_{\rm BH}\sim 10^6M_{\odot}$), which depends on the slope of the power-law continuum spectrum of the corona}. It is clear that continuum emission from accretion onto the first massive black holes can be easily detected with future, and even existing telescopes (though in detail this depends on $\alpha_{X}$, and where this power-law spectrum cuts off). The more pressing question is therefore how we would be able to identify these objects. \\

Newly formed massive black holes are thought to rapidly acquire an accompanying stellar component, mostly from mergers with the nearby star forming halos which provided the required Lyman-Werner flux\cite{Dijkstra08,Agarwal12,Regan17} to enable direct collapse of a primordial gas cloud into a massive black hole\cite{Agarwal16,Hartwig16}. The mass of the black hole is initially large compared to its accompanying stellar mass, and these galaxies - at least initially - do not obey the locally observed relation that the central black hole mass of a galaxy corresponds to $\sim 0.1\%$ of its stellar mass (in the bulge). This initial stage - during which the black hole is overly massive - has been referred to as the `Obese Black Hole Galaxy stage (OBG stage, see Ref~\refcite{Agarwal13}). Ref~\refcite{Priya17} provide several observational diagnostics which may help us identify galaxies in their OBG stage, including: ({\it i}) a large bolometric luminosity ($L_{\rm bol} \gsim 10^{44}$ erg s$^{-1}$, which corresponds to the Eddington luminosity of a black hole of mass $M_{\rm BH} \sim 10^6 M_{\odot}$), ({\it ii}) a large ratio of X-ray to optical flux, and ({\it iii}) particular colors in the JWST bands. Additional ways to identify galaxies hosting newly formed massive black holes may be through line emission, as we discuss next.

\subsection{Direct Detection of Lines from the First Massive BHs}\label{sec:line}

Reprocessing of ionizing radiation in the hydrogen gas surrounding the accreting black hole gives rise to strong recombination lines. High-energy UV ($E \gsim 55.4$ eV) and X-ray photons can doubly ionize Helium. Recombining doubly ionized Helium produces strong He1640 line emission. The strongest line that is produced however, is H Ly$\alpha$, which can account for as much as $\sim 40\%$ of the bolometric luminosity of the accreting black hole (see e.g. Ref~\refcite{Sobral16}). \\

The recombination lines contain large fluxes, but condensed into narrow spectral ranges, which can make them more easily detectable. For example, a Ly$\alpha$ line with a luminosity of $L_{\alpha}\sim 10^{43}-10^{44}$ erg s$^{-1}$ can be detected out to $z\sim 7$ with existing large telescopes such as VLT, Keck \& Subaru. At higher redshifts, the Ly$\alpha$ line redshifts into the NIR where the line is better accessible from space, as atmospheric OH lines contaminate ground-based observations. NIRSPEC on JWST has the sensitivity to detect line fluxes of $\sim 1-2\times 10^{-18}$ erg s$^{-1}$ cm$^{-2}$ at $\lambda \sim 1-5 \mu$m, which corresponds to (apparent) Ly$\alpha$ luminosities of $\L_{\alpha} \sim 10^{42}$ erg s$^{-1}$ at $z\sim 10$. JWST should therefore be able to access Ly$\alpha$ line emission even from the lower mass `massive' seeds ($M_{\rm BH} \sim 10^4 M_{\odot}$). For comparison, for `moderately extended sources', the 7$\sigma$ detection threshold for emission lines in WFIRST is $\sim 1-1.5\times 10^{-16}$ erg s$^{-1}$ cm$^{-2}$ at $\lambda \sim 1-2 \mu$m (see Fig~1 of \citet{Gehrels15}), which corresponds to (apparent) Ly$\alpha$ luminosities of $\L_{\alpha} \sim 10^{44}$ erg s$^{-1}$ at $z\sim 10$. EUCLIDs deep, slitless spectroscopic surveys are expected to go similarly deep to $\sim 10^{-16}$ erg s$^{-1}$ cm$^{-2}$ at comparable wavelengths (see e.g. \citet{Laureijs2011}).  WFIRST \& EUCLID may thus be able to directly detect line emission from the most massive black hole seeds. What is more important is that EUCLID \& WFIRST will provide interesting targets which can then be followed-up spectroscopically with e.g. JWST and the next generation of ground-based facilities such as E-ELT. The E-ELT is expected to be able to detect similar line fluxes to $\lambda \lsim 2.5\mu$m, by suppressing the narrow atmospheric OH lines. Because JWST and future ground-based facilities are expected to operate deeper into the IR, they can access other redshifted hydrogen lines such as the H$\alpha$ line is expected to be a factor of $\sim 8$ times weaker when powered by recombination. \\

There has been some speculation that we may have already detected such a newly formed massive black hole: CR7 (Cosmos-Redshift 7, \citet{Sobral}) is one of the most luminous ($L_{\alpha} \sim 10^{44}$ erg s$^{-1}$) Ly$\alpha$ emitting galaxies in the Universe, and at $z\sim 6.6$. In addition, the Ly$\alpha$ line is unusually strong, with a rest-frame equivalent width of EW$_{\alpha}\sim 200$ \AA. The strength of this line is especially surprising given that the Ly$\alpha$ forest\footnote{The precise opacity of the IGM to radiation that emerges from a galaxy close to the Ly$\alpha$ resonance is strongly frequency-dependent. How much flux is transmitted through the IGM at $z\sim 6.6$ depends on the spectral shape and shift of the Ly$\alpha$ line as it emerges from the galaxy (see e.g. \cite{Review} for an extended review)} is opaque to Ly$\alpha$ emission. In addition, CR7 appears to have a strong He1640 line, which is expected if CR7 harbors a source of hard ionizing radiation, such as an accreting black hole. CR7 further consists of several components, one of which consists of an older, passive galaxy. This has fueled speculation that CR7 may harbor a black hole that formed via direct collapse\footnote{Other explanations for CR7 have been put forward. Strong Ly$\alpha$, He1640 line emission, combined with the blue continuum of one of the three components of CR7 has also been associated with Population III star formation (see e.g. \citet{Sobral,Visbal16}). However, recent theoretical studies indicate that the required mass of Population III mass that is required to explain CR7 exceeds the maximum mass that is considered theoretically reasonable \citep{Visbal17,Xu16}. Recent observations by \citet{Bowler17} imply that the strength of the He1640 is weaker, and that [OIII] may have been detected, which would rule out the Population III scenario (but not the DCBH-scenario, see  \citet{Pacucci17}).}, as the older galaxy would have been a luminous source of Lyman-Werner flux radiation - a key ingredient for direct collapse - when it was forming stars in its past \citep{Agarwal16,Hartwig16}.\\

The extreme physical conditions of the HI gas surrounding the accreting black holes (high density, metal \& $H_2$-free environments) may give rise to interesting Ly$\alpha$ radiative transfer effects, which can give rise to unusual line ratios (this needs to be explored in future work, also see \S~\ref{sec:formation}). What is in this sense peculiar about CR7 is that the spectral shape of the Ly$\alpha$ line is similar to that of low-redshift strong Ly$\alpha$ emitting galaxies, in which the Ly$\alpha$ spectral line shape is affected by (likely stellar feedback-driven) outflows. This is important for two reasons: ({\it i}) the interstellar medium of CR7 is likely shaped by processes similar to that in the local Universe. This is consistent with a scenario in which the dark matter halo hosting a direct collapse black hole rapidly merges with the star forming halo (as we described when introducing the OBG stage). This merging process likely re-shapes the physical conditions of the interstellar medium in a way that more closely resembles that of `ordinary' galaxies ({\it ii}) it implies that the accreting black hole is {\it not} surrounded by a large column density of HI gas, as is thought to be the case for newly formed DCBHs. This is because large column densities of HI gas require Ly$\alpha$ photons to scatter frequently before escaping, which broadens the width of the spectral line. The fact that the observed Ly$\alpha$ line is narrow, limits the allowed HI column densities through which Ly$\alpha$ photons are scattering to be $\sim 10^{19}-10^{20}$ cm$^{-2}$ (see Ref~\refcite{Sobral16}).

\subsection{Detection of the Formation of the First Massive BHs}\label{sec:formation}

Ref~\refcite{Kohei} discuss observational signatures of a particular formation channel of massive ($M_{\rm BH}\sim 10^5 M_{\odot}$) black holes via Hyper-Eddington accretion. In this scenario, the mass accretion rate onto a central black hole (which could be stellar mass and formed from a Pop III star, or more massive and formed differently) exceeds $\dot{M}_{\rm BH}\gsim 5000 L_{\rm edd}/c^2$. The transition to hyper-Eddington happens when the radius HII region created by hot gas accreting onto the black hole becomes smaller than the Bondi radius of the black hole. Under these conditions, photons are trapped in the accretion flow and carried into the black hole together with the gas. The overall bolometric luminosity of the accreting black hole is the Eddington luminosity. The most massive of these black holes ($M_{\rm BH} \sim 10^6 M_{\odot}$) could also be accompanied with a Ly$\alpha$ luminosity reaching $\sim 10^{42}-10^{44}$ erg s$^{-1}$. Given the lower gas temperatures involved in this mechanism, the He1640 line should be absent. The spectral shape of the Ly$\alpha$ line could distinguish it from other models, though this requires Ly$\alpha$ radiative transfer modelling which needs to be explored in future work. Ref~\refcite{Kohei} further argue that this Ly$\alpha$ signature likely disappears in the presence of small amounts of dust, which is interesting as it would suggest it is uniquely associated with the {\it first} black holes that form this way.\\

\begin{figure}
\centerline{\includegraphics[width=12cm]{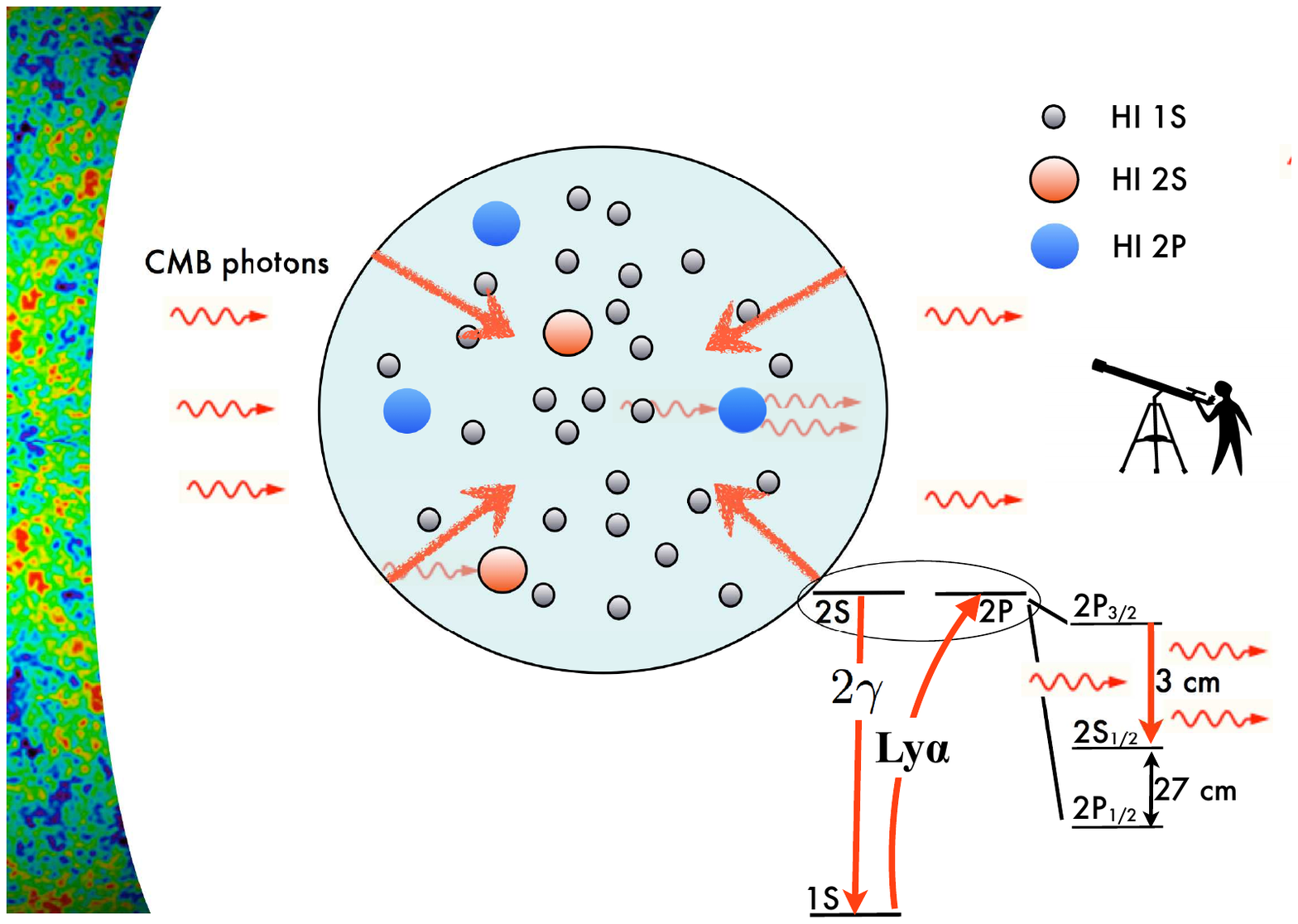}}
\caption{The physical conditions inside a gas cloud that collapses directly into a massive black hole are optimal for trapping Ly$\alpha$ radiation. Trapped Ly$\alpha$ `pump' the $2p$-level of HI, and can give rise to stimulated fine-structure emission at a rest-frame wavelength of $\lambda=3.04$ cm. Stimulated 3-cm emission would be a smoking gun signature of direct collapse {\it Credit: from Figure 1 of \citet{Shiv}.  \textcopyright AAS. Reproduced with permission.}}
\label{fig:3cmmaser}
\end{figure}
Another unique signature that is associated with the formation of direct collapse black holes was discussed in Ref~\refcite{Shiv}: the physical conditions that enable direct collapse: ({\it i}) the collapse of pristine gas, ({\it ii}) the (virtual) absence of H$_2$, ({\it iii}) the large column densities of HI gas, and ({\it iv}) suppressed fragmentation, provide optimal conditions for Ly$\alpha$ trapping, which can cause Ly$\alpha$ photons to scatter $\gsim 10^8-10^9$ times. Under these conditions the $2p$ level (which is the only level of atomic Hydrogen excited by Ly$\alpha$ absorption) can be overpopulated with respect to the $2s$ level. Spin-orbit coupling and the Lamb-shift causes one the $2p_{3/2}$ ($2p_{1/2}$) level\footnote{We adopt the notation $nL_{J}$, where $n$ is the principle quantum number, $L$ denotes the electron's {\it orbital} angular momentum, and $J$ denotes the {\it total} (orbital + spin) quantum number.} to lie above (below) the $2s$ level, and give rise to a fine-structure line with rest-frame wavelength $\lambda=3.04$ cm ($\lambda=27$ cm). Overpopulation of the $2p$ level can therefore give rise to stimulated 3-cm emission\cite{Field61}. Ref~\refcite{Shiv} showed that gas clouds directly collapsing into a massive black holes could amplify the background CMB (see Fig~\ref{fig:3cmmaser}) at a level that is detectable with SKA1-mid. This signal is only detectable during a small fraction of the collapse of the cloud, and beaming effects which occur in saturated masers may also reduce the detectability of  the signal. However, the 3-cm line has a broad peculiar spectral shape, which makes it easy to distinguish from other lines. If detected, this would provide a true smoking-gun signature of direct collapse.

\section{Conclusions \& Outlook}\label{sec:conc}

The main goal of this chapter is summarized in Figure~\ref{fig:summary}: we have presented a brief discussion of the prospects for detecting the first generations of black holes in our Universe. To summarize:
\begin{figure} 
\centerline{\includegraphics[width=12cm]{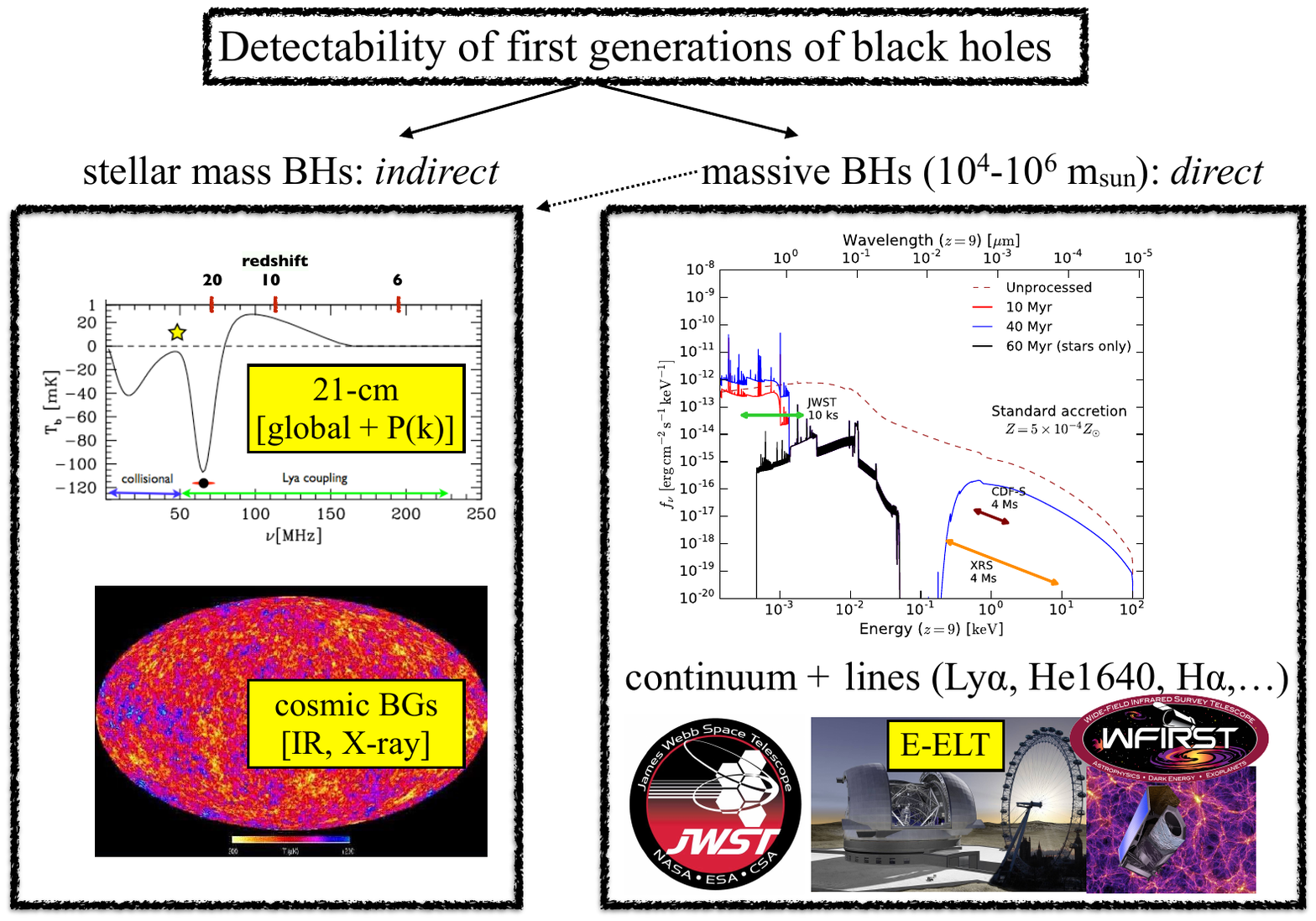}}
\caption{Schematic representation of this chapter: we have discussed ({\it i}) the {\it indirect} detectability of the first generations of stellar mass black holes through their impact on various cosmic radiation backgrounds, ranging from the 21-cm background ({\it top-left}) to the IR and X-ray backgrounds ({\it lower left}), ({\it ii}) the direct detectability of continuum and line emission from the first more massive black hole seeds with a range of future space and ground-based telescopes such as WFIRST, EUCLID, JWST, and E-ELT ({\it right}). }
\label{fig:summary}
\end{figure}
\begin{itemize}[leftmargin=*]
\item The first stellar mass black holes can only be detected indirectly: they leave a unique imprint on the 21-cm background in several ways: ({\it i}) the presence and/or depth of the global 21-cm absorption trough is directly linked to the emergence of the first black holes and their spectral properties; ({\it ii}) X-ray heating by the first black holes directly affects the spatial fluctuations of the 21-cm signal in ways that can be detected at $\sim$ hundreds-$\sigma$ level by future low-frequency arrays such that SKA1-low and HERA. Early generations of stellar mass black holes may have contributed significantly to the unresolved soft-XRB and the near-IR background.\\

\item If the first black holes were more massive, as has been argued theoretically, then this greatly improves the prospects for detecting them. Their continuum emission is well within reach of future facilities operating in the near-IR such as WFIRST, JWST, large ground-based facilities (E-ELT, GMT and TMT), while their X-ray emission is within reach of the deepest X-ray observations that have been performed with Chandra. Future X-ray telescopes (ATHENA, Lynx) should be able to detect X-ray emission from individual high-z massive black holes. It is theoretically possible to identify these black holes among other 'normal' high-z galaxies through a combination of the ratio of their optical-to-X-ray emission, and through their broad-band colors. Recombination lines provide another promising way of both finding and identifying the most distant, massive newly formed black holes, though quantitative predictions require further radiative transfer modelling.
\end{itemize}

In short: the next generation of telescopes will clearly have the capabilities to detect (indirectly and/or directly) electromagnetic radiation from the first generations of black holes. These detections (and/or non-detections) will provide the first observational constraints on the formation of the first baryonic structures in our Universe, which will represent a milestone for observational cosmology.

\section*{Acknowledgements}
%\section{Acknowledgments}
I would like to thank Marat Gilfanov, Felix Mirabel, and Philippe Laurent for helpful correspondence, Andrei Mesinger for helpful correspondence and for permission to use Fig~2, and Priya Natarajan for permission to use Fig~3. I would like to thank Greg Salvesen and Mike McCourt for helpful discussions, and the physics department at UCSB for their kindly hosting me while I wrote this chapter.

\bibliographystyle{ws-rv-van}
%\bibliography{bhrefs}

%\blankpage
\printindex[aindx]                 % to print author index
\printindex                         % to print subject index

\end{document}